\PassOptionsToPackage{dvipsnames}{xcolor}
\documentclass[sigconf,nonacm]{acmart} 

\AtBeginDocument{%
  }

\usepackage{algorithm}
\usepackage{algpseudocode}
\usepackage{tabularx}
\usepackage{xcolor}
\usepackage[most]{tcolorbox}
\usepackage{booktabs}
\usepackage{multirow}
\usepackage{multicol}
\usepackage{graphicx}
\usepackage{wrapfig}
\usepackage{xspace}
\usepackage{enumitem}
\usepackage{listings}
\usepackage[utf8]{inputenc}
\usepackage[T1]{fontenc}
\usepackage{microtype}
\usepackage{setspace,caption}
\usepackage{subcaption}
\usepackage{flushend}
\usepackage{url}
\usepackage{hyperref}
\usepackage{cleveref}

\definecolor{KWColor}{rgb}{0.37,0.08,0.25}
\definecolor{CommentColor}{rgb}{0.12,0.38,0.18}
\definecolor{StringColor}{rgb}{0.06,0.10,0.98}
\definecolor{darkred}{rgb}{0.75,0,0}
\definecolor{lightgrey}{rgb}{0.8,0.8,0.8}

\lstdefinestyle{Eclipse}{
  xleftmargin=0pt,
  basicstyle=\ttfamily\footnotesize,
  commentstyle=\color{CommentColor}\ttfamily\footnotesize,
  stringstyle=\color{StringColor},
  keywordstyle=\color{KWColor}\bfseries,
  escapeinside={/*@}{@*/}
}

\crefname{figure}{Figure}{Figures}
\crefname{table}{Table}{Tables}
\crefname{section}{Section}{Sections}
\crefname{equation}{Equation}{Equations}
\crefname{algorithm}{Algorithm}{Algorithms}
\crefname{lstlisting}{Listing}{Listings}

\begin{document}

\title{FlowETL: An Autonomous Example-Driven Pipeline for Data Engineering}

\author{Mattia Di Profio}
\email{m.diprofio.21@abdn.ac.uk}
\affiliation{%
  \institution{University of Aberdeen}
  \department{Department of Computing Science}
  \city{Aberdeen}
  \country{UK}
}

\author{Mingjun Zhong}
\email{mingjun.zhong@abdn.ac.uk}
\affiliation{%
  \institution{University of Aberdeen}
  \department{Department of Computing Science}
  \city{Aberdeen}
  \country{UK}
}

\author{Yaji Sripada}
\email{yaji.sripada@abdn.ac.uk}
\affiliation{%
  \institution{University of Aberdeen}
  \department{Department of Computing Science}
  \city{Aberdeen}
  \country{UK}
}

\author{Marcel Jaspars}
\email{m.jaspars@abdn.ac.uk}
\affiliation{%
  \institution{University of Aberdeen}
  \department{Department of Chemistry}
  \city{Aberdeen}
  \country{UK}
}

\begin{abstract}
The Extract, Transform, Load (ETL) workflow is fundamental for populating and maintaining data warehouses and other data stores accessed by analysts for downstream tasks. A major shortcoming of modern ETL solutions is the extensive need for a human-in-the-loop, required to design and implement context-specific, and often non-generalisable transformations. While related work in the field of ETL automation shows promising progress, there is a lack of solutions capable of automatically designing and applying these transformations. We present FlowETL, a novel example-based autonomous ETL pipeline architecture designed to automatically standardise and prepare input datasets according to a concise, user-defined target dataset. FlowETL is an ecosystem of components which interact together to achieve the desired outcome. A Planning Engine uses a paired input-output datasets sample to construct a transformation plan, which is then applied by an ETL worker to the source dataset. Monitoring and logging provide observability throughout the entire pipeline. The results show promising generalisation capabilities across 14 datasets of various domains, file structures, and file sizes.
\end{abstract}


\maketitle

\section{Introduction}\label{sec:intro}

With the increasing prevalence of Big Data, the demand for robust data engineering solutions has grown significantly. Data engineering, otherwise known as data wrangling, involves collecting, organising, and transforming raw data into a format suitable for downstream tasks~\cite{furche}. A common implementation of these tasks comes in the form of an Extract, Transform, Load (ETL) pipeline, where data is extracted from multiple sources, transformed into a desirable format, and loaded into its final destination for further consumption. The literature estimates that around 80\% of a data analyst's time is spent on data wrangling tasks due to absence of a reliable and efficient procedure to transform data automatically~\cite{mbata}.

Various ETL tools have been developed in the literature~\cite{jin,foofah,devarasetty,bodensohn,Sharma,Lip}. Such tools provide rich recipes, from intuitive methodologies to Large Language Model (LLM) approaches, for wrangling various types of data. However, using these tools still requires substantial manual effort, and the main challenge lies in achieving an autonomous and reliable data engineering approach. In other words, there is a persistent need for a framework which minimises the data scientists' efforts while providing a highly resilient and adaptable solution to data engineering~\cite{dakrory}.

To address these challenges, this paper presents FlowETL: a fully autonomous, example-based ETL pipeline architecture designed to minimise developer effort while ensuring robustness and adaptability. FlowETL can automatically handle common data quality issues such as missing values, numerical outliers, data canonicalization (standardisation), and data de-duplication~\cite{nazabal}. Specifically, for data canonicalization, our framework performs schema and instance-level standardisation using only a user-provided pair of sample input-output files from which desired transformations are inferred. Our main contribution with this work is that we have developed an ETL framework capable of addressing the data preparation needs of a wide range of downstream tasks, reducing human involvement in ETL workflow design and monitoring, and maintaining high performance and output accuracy. Additionally, this research explores the integration of LLMs into the ETL process, contributing to the emerging field of automated data pipelines construction.

In summary, this paper's contributions are:
\begin{itemize}
  \item We present FlowETL, a novel autonomous ETL pipeline architecture that leverages example-driven transformation inference to minimize manual intervention.
  \item We introduce a comprehensive system design including Planning Engine, ETL Worker, and monitoring components that work together to achieve end-to-end automation.
  \item We evaluate our approach on 14 diverse datasets from various domains, demonstrating robust generalisation capabilities and high data quality retention.
  \item We provide a detailed comparison with existing ETL tools, showing the effectiveness of our autonomous approach compared to manual pipeline construction.
\end{itemize}

The remainder of this paper is structured as follows: \cref{sec:related} reviews relevant background and related work. \cref{sec:architecture} describes the design of the FlowETL system. \cref{sec:evaluation} presents the evaluation methodology and experimental results. \cref{sec:conclusion} discusses the system's strengths, limitations, and potential directions for future work.

\section{Related Work}\label{sec:related}

This section introduces the Extract, Transform, Load (ETL) workflow and its autonomous variants, with a focus on pattern-driven and example-driven approaches. Particular attention is given to the schema matching stage, occurring within the transformation step, for which a brief taxonomy of techniques proposed in the literature is presented.

\subsection{The ETL Workflow}

The Extract, Transform, Load (ETL) workflow is commonly used to populate data warehouses. In the extraction phase, heterogeneous data sources are ingested into a staging area to facilitate standardisation. The transformation step focuses on addressing issues such as missing values, duplicates, and format inconsistencies. This stage encompasses data cleansing, normalisation, and enrichment, all of which promote high data quality and optimise the performance of downstream tasks~\cite{vassiliadis}. Finally, the loading phase delivers the transformed data to target systems for consumption. ETL tools are typically categorised as graphical-based, which offer intuitive interfaces for less technical users, and scripting-based, which provide greater flexibility at the cost of increased complexity.

\subsection{Autonomous ETL Pipelines}

Recent advances in ETL pipeline development have focused on automating the transformation stage to mitigate against the overhead associated with designing transformation and enrichment workflows.

Mondal et al.~\cite{mondal} propose a system wherein a Recommender consisting of multiple machine learning algorithms interacts with a Reporting Agent to identify performance bottlenecks and optimise the overall pipeline efficiency. Devarasetty~\cite{devarasetty} argues that traditional ETL pipelines are often vulnerable to changing requirements, and leveraging machine learning could enable dynamic workflow adjustments while reducing manual effort and improving data quality. The proposed solution uses Random Forests and Auto-encoders for anomaly detection, Natural Language Processing for data tagging, and Recurrent Neural Networks (RNNs) to automatically infer and apply data transformations.

Recent advancements in large language models (LLMs) have led to the emergence of autonomous pipeline architectures. Bodensohn et al.~\cite{bodensohn} evaluate the use of LLMs for fully autonomous data wrangling, identifying high operational costs and limited generalisation to enterprise data as critical barriers to practical adoption.

\subsection{Pattern-Driven Architectures}

Jin et al.~\cite{jin} introduce a pattern-based approach, Transform-by-Pattern (TBP), tailored towards data repair and integration. TBP represents transformations as a triplet $(P_s, P_t,T)$, where $P_s$ and $P_t$ are syntactic regex patterns describing the source and target columns for which the program $T(P_t, P_s)$ is applicable. In the TBP approach, transformation programs are inferred using a pre-constructed lookup table. Despite promising results, the method is limited by the high computational cost and time required to build this table from large-scale public data sources.

\subsection{Example-Driven Architectures}

Foofah~\cite{foofah} standardises the structure of input datasets by inferring a transformation plan from a corresponding target dataset. It models transformation inference as a state-space search problem, using an A*~\cite{hart} heuristic search with pruning strategies to improve efficiency while searching for a suitable transformation plan. While Foofah successfully synthesised correct programs in 90\% of test cases during its evaluation phase, the limited handling of instance-level transformations restricts its applicability to more complex data engineering scenarios.

DataXFormer~\cite{morcos} performs transformation inference by using user-provided input and output column labels to query and extract relevant data from a large corpus of web tables and forms. Unlike example-driven systems, it relies on column headers to construct keyword queries. While effective for structured sources, DataXFormer struggles with unstructured data and requires substantial setup, utilising approximately 112 million web forms and tables during its search phase.

He et al.~\cite{heTBE} propose Transform-by-Example (TBE): a system that synthesises transformation programs suitable for the source-target dataset pair provided by searching an index of over 50,000 functions sourced from code-sharing platforms. This approach supports a wide range of syntactic and semantic transformations. However, its effectiveness is limited in under-represented domains due to the scarcity of relevant functions within the index.

In more recent work, Singh et al.~\cite{singh} developed a system which uses a domain-specific language for semantic string transformations and syntactic operations. It works by taking input-output examples and searching through a space of candidate transformations and using a ranking heuristic to select the most suitable one. As users provide more examples, the system iteratively refines its output, making it a semi-autonomous solution. A major limitation is the size of the possible transformations set, which can be extremely large even for simple examples, posing scalability issues. Additionally, the system struggles to handle infinite domains such as arithmetic transformations on numerical instances.

\subsection{Schema Matching}

Schema matching is a fundamental step of dataset standardisation and is often integrated into broader ETL workflows~\cite{bernstein}. It involves identifying meaningful correspondences between columns in a source and target dataset. Across the literature, the notion of a \emph{Match} operator is mentioned extensively, with Rahm et al.~\cite{rahm} defining it as a function which produces a mapping between elements of the two schemas that semantically correspond to each other, and Giunchiglia et al.~\cite{giunchiglia} describing it as an operator which produces a mapping between nodes of a bipartite graph. Formally, the schema matching problem can be modelled as an undirected bipartite graph $G = (X,Y,E)$, where $X$ and $Y$ represent the source and target columns respectively and $E$ represent weighted edges between $X$ and $Y$, with each edge $(x, y) \in E$ having a weight $w \in [0, 1]$ based on a similarity score between $x \in X$ and $y \in Y$. Building upon this abstraction, a schema match $M$ is then defined as a subset of $E$, classified by Gal et al.~\cite{gal} as one of the following:
\begin{itemize}
    \item $1:1$, where $M$ is a subset of pair-wise disjoint edges.
    \item $1:n$, where $x \in X$ maps to multiple $y_1,y_2,\dots \in Y$, known as Replication.
    \item $n:1$, where multiple $x_1,x_2,\dots \in X$ map to a single $y \in Y$, known as Decomposition.
\end{itemize}

\subsection{Overview of Schema Matching Approaches}

Bernstein et al.~\cite{bernstein} emphasise the limitations of one-shot schema matchers, which may yield suboptimal or incorrect results due to misalignment with user intent and a high incidence of false positives requiring manual correction. To address this, they propose an iterative ranking approach that combines schema-based heuristics with user interaction history. The system progressively refines matches based on previous user-validated mappings. However, its suitability for larger schemas is limited due to a linear increase in user effort with the size of the matching problem associated with the use a graphical interface. Gal et al.~\cite{gal} address the limitations of one-shot schema matchers by simultaneously generating multiple schema mappings and evaluating them using a \emph{Stability Heuristic}, which positively scores mappings that consistently appear across the top $K$ generated candidates. This approach supports $1:1$, $1:n$, and $n:1$ mappings. However, it is computationally intensive for large datasets and relies heavily on the quality of initial mappings.

MAXSM~\cite{beg} proposes a heuristic-based approach to XML schema matching, which incrementally computes similarities between schema elements using a similarity function $sim(x,y) \rightarrow [0,1]$ computed by combining multiple heuristics, such as natural language similarity through WordNet, a tree-spanning method to detect structurally similar node clusters, and location path-based heuristics to assess node similarity. It then applies threshold-based decision-making to determine potential mappings. These similarity scores are recorded in a matrix, which informs the final mapping decisions in the output construction phase. While MAXSM poses a promising multi-heuristic approach, it remains unimplemented and has not yet undergone empirical evaluation.

ReMatch~\cite{sheetrit} leverages retrieval-enhanced large language models (LLMs) to assist human schema matchers without requiring predefined mappings, training, or direct access to source databases. It converts target tables and source attributes into structured documents, applies text embeddings to retrieve top candidate tables based on semantic similarity, and employs an LLM to rank the most similar target attributes, generating a ranked list of potential matches.

Magneto~\cite{liu} employs a two-stage architecture consisting of a lightweight candidate retriever followed by a more computationally intensive re-ranker. This design aims to reduce schema matching costs while preserving high accuracy. Although results are promising, the system was evaluated exclusively on medical domain data, limiting insight into its generalisability across other domains.

In summary, a major shortcoming of modern ETL solutions is the extensive need for a human-in-the-loop to design and implement context-specific, often non-generalisable transformations. While related work in ETL automation shows promising progress, there is still a lack of solutions capable of automatically designing and applying these transformations without relying on large underlying datasets of transformation examples. To account for these shortcomings, we present FlowETL: an example-based autonomous ETL pipeline architecture designed to automatically standardise and prepare input datasets according to a concise, user-defined target dataset.

\section{System Architecture}\label{sec:architecture}

\begin{figure}[t]
    \centering
    \includegraphics[width=0.4\textwidth]{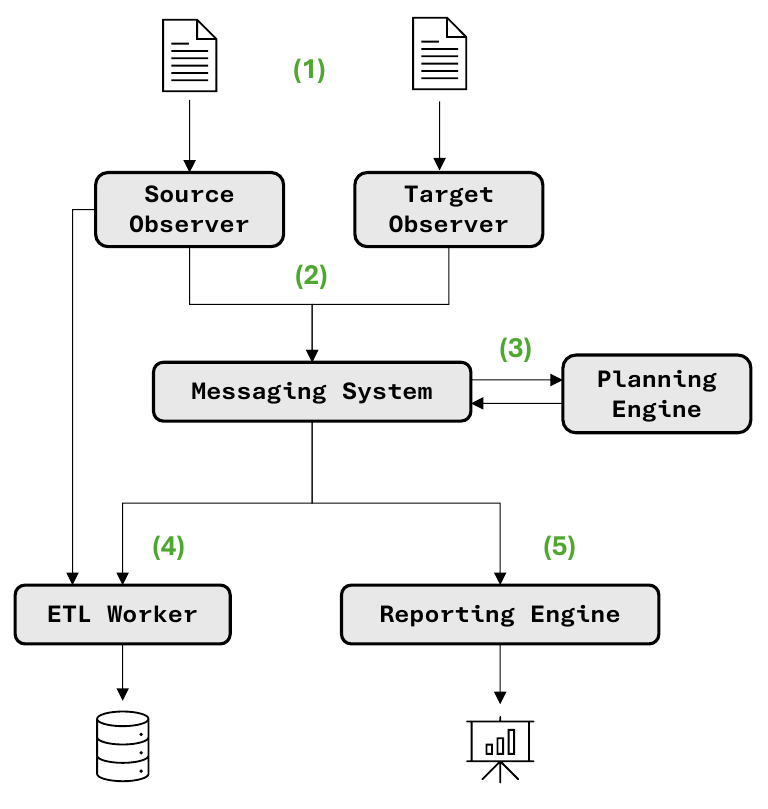}
    \caption{Simplified FlowETL system architecture. Observers detect source and target files (1), then upload them to the messaging system (2). The Planning Engine consumes them, computes and publishes a plan (3) which is then applied to the source file by the ETL Worker (4) and loaded elsewhere. The Reporting Engine gathers runtime metrics and produces a report (5).}
    \label{fig:simplified_architecture}
\end{figure}

In this section we introduce the main components of the architecture, their role, and their interactions. As shown in \cref{fig:simplified_architecture}, the Observers detect the source and target datasets for a particular source. Using these datasets, the Planning Engine synthesises a plan which both standardises the source file according to the target and improves its estimated data quality with respect to missing values, duplicates, and outliers. The ETL Worker then applies the plan to the entire source dataset. The Messaging System and Report Engine enable sharing of payload and information between components and logging \& reporting of runtime metrics, respectively.

\subsection{Abstractions}

To enable support for both structured and semi-structured datasets, two core abstractions have been made. The first defines a set of internal data types, while the second abstraction translates both structured and semi-structured files into an Internal Representation (IR), providing an interface each component can utilise to operate on the source file's contents.

\subsubsection{FlowETL Data Types System}

We define a set of generalised data types capable of modelling a wide range of heterogenous data points, deliberately limiting the edge cases encountered during type conversion. To ensure comprehensive representation of data in both structured (CSV) and semi-structured (JSON) datasets, the internal data types system should support primitive types (e.g., string, float, boolean) as well as complex nested structures such as lists and objects, commonly found in datasets containing JSON data. While existing data models from Pandas~\cite{pandas} and Apache Spark~\cite{apachespark} were considered for their operational advantages, custom data types were ultimately developed to maximise flexibility and avoid dependency constraints imposed by these frameworks. The FlowETL data types and corresponding abstracted types are outlined in \cref{tab:flowetl_types}.

\begin{table}[t]
\centering
\caption{Column data types supported internally by FlowETL and the corresponding data types abstracted.}
\label{tab:flowetl_types}
\begin{tabular}{| c | l |}
\hline
\textbf{Data Type} & \textbf{Description} \\
\hline
\texttt{number} & Numerical values \\
\hline
\texttt{string} & Characters and strings \\
\hline
\texttt{boolean} & Symbols representing truth values \\
\hline
\texttt{complex} & Lists, dictionaries \\
\hline
\texttt{ambiguous} & Columns containing multiple data types \\
\hline
\end{tabular}
\end{table}

\subsubsection{Internal Representation (IR)}

Abstracting both structured and unstructured files into an IR allows for a pipeline architecture which can easily be adapted to handle other file types, such as XML, by simply providing the conversion logic to translate the file into an IR. Moreover, this abstraction allows for increased maintainability. For instance, each Data Task Node (DTN) can be defined only once and designed to operate on the IR, therefore there is no need for a separate DTN variant to handle each of the supported file types. The file-to-IR translation mechanism works by extending unstructured files into a tabular format inspired by Dataframes, a construct utilised in both Apache Spark~\cite{apachespark} and Pandas~\cite{pandas}. For CSV files, the algorithm is trivial: the file's contents are translated into a 2-D matrix with the first row denoting the column names and each subsequent row representing a row in the original file. For JSON files, the process is more complex. A major assumption made during the read-in process is that the source data within the JSON file is stored in a list of JSON objects. The extraction logic recursively traverses the possibly nested structure of a JSON file until a value of type \texttt{list} is found and returned.

\begin{figure}[t]
    \centering
    \begin{minipage}{0.45\textwidth}
        \centering
        \begin{lstlisting}[language=Python, basicstyle=\footnotesize\ttfamily]
   {objects: [
     {ID: 1, name: John, age: 50, salary: 1234}, 
     {ID: 2, name: Amy, salary: 5678},
     {ID: 3, name: Ellie}
   ]}
        \end{lstlisting}
        \vspace{-1em}
    \end{minipage}\hfill
    \vspace{1em}
    \begin{minipage}{0.45\textwidth}
        \centering
    \begin{tabular}{|l|l|l|l|}
        \hline
        ID & name & age & salary \\
        \hline
         1 & John & 50 & 1234 \\
        \hline
         2 & Amy & * & 5678 \\
        \hline
         3 & Ellie & * & * \\
        \hline
    \end{tabular}
        \vspace{1em}
    \end{minipage}
    \caption{JSON payload with associated \texttt{objects} reconstruction key (top) and its internal representation within FlowETL (bottom).}
    \label{fig:translation_mechanism}
\end{figure}

Subsequently, a union of keys from all objects in the list is created to form the headers of the IR. Each object's keys are extended to include all the headers, with placeholder values added for any extended keys. The values for each object are then translated into rows in the IR, serialising complex objects such as dictionaries and lists. The translation process from IR back to the source dataset involves creating a dictionary for each row, where column names map to keys and cell contents are assigned as values, excluding any placeholders. This translation mechanism, shown in \cref{fig:translation_mechanism} avoids any loss of information when translating to the IR or gain of redundant information when translating back up to the original file type.

\subsection{Source and Target Observers}

The Observers are designed to prepare the files of interest through a series of pre-processing steps. Namely, these steps are ingesting the source-target files pair, translating them to an Internal Representation (IR), and publishing these artifacts to the Messaging System. Both observers additionally record information at runtime about the files processed, such as file name, objects count, file size, and an optional key returned by the inward translation mechanism when handling JSON files, as shown in \cref{fig:translation_mechanism}. These metrics, shown in \cref{fig:observers_payload}, are separately published to a dedicated topic on the Messaging System to be then consumed by the Reporting Engine throughout the pipeline runtime.

\begin{figure}[t]
    \centering
    \begin{minipage}{0.4\textwidth}
        \centering
        \begin{lstlisting}[language=Python, basicstyle=\footnotesize\ttfamily]
        { "from": observerName, 
          "contents": { 
            "filename": filename, 
            "objectsCount": count, 
            "filesizeMBs": fileSize
        }}
        \end{lstlisting}
        \vspace{-1em}
        \caption*{(a) Runtime metrics payload}
    \end{minipage}\hfill
    \begin{minipage}{0.4\textwidth}
        \centering
        \begin{lstlisting}[language=Python, basicstyle=\footnotesize\ttfamily]
        {
            "name": filename, 
            "reconstructionKey": key, 
            "contents": sampledIR
        }
        \end{lstlisting}
        \vspace{-1em}
        \caption*{(b) Source file artifacts payload}
    \end{minipage}
    \caption{Structure of Observer payloads sent to the Messaging System.}
    \label{fig:observers_payload}
\end{figure}

When the Source Observer is triggered, it randomly samples a subset of rows from the IR. This step is carried out to reduce latency and processing time, especially during the plan construction phase. Although this approach is expected to facilitate faster information exchange between components, it may compromise the quality of the computed plan, as it relies on the assumption that the data within the source dataset is randomly distributed. When the Target Observer is triggered, no sampling is performed, since any target file is expected to be relatively small (5-20 objects at most).

\subsection{Messaging System}

\begin{figure}[t]
    \centering
    \includegraphics[width=0.4\textwidth]{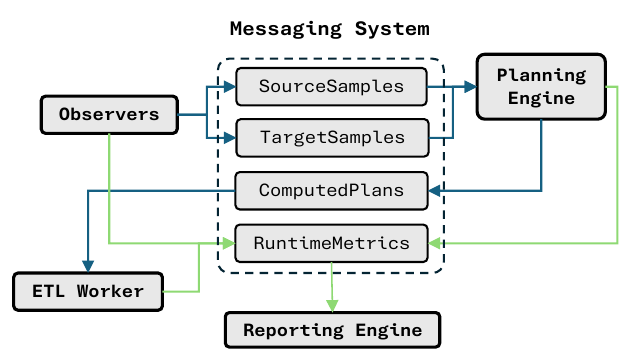}
    \caption{Messaging system topics overview. Blue edges denote file data, green edges denote runtime metrics. Incoming edges denote data being published to a topic, outgoing edges denote data being consumed from a topic.}
    \label{fig:kafka_broker}
\end{figure}

Several key requirements were identified for a robust messaging system. Namely, isolation of communication between components without exposing messages to non-participating parties, scalability to meet system demands, and support for sequential message processing from specific offsets to prevent bottlenecks, particularly as the Planning Engine may be overwhelmed if the Observers publish artifacts faster than they can be processed. Apache Kafka~\cite{kafka} was selected as a suitable solution due to its implementation of the Publisher-Subscriber model~\cite{pubsub}, which enables decoupled and anonymous communication via topic-based message exchange. A critical risk noted was the system's reliance on the messaging infrastructure, making component communication vulnerable during failures. Consequently, strong availability and reliability guarantees were required. Kafka addresses these concerns through replication and fault tolerance mechanisms~\cite{kafka_guarantees}, significantly reducing failure risks. An overview of the Messaging System is shown in \cref{fig:kafka_broker}.

\subsection{Planning Engine - Overview}

The Planning Engine is responsible for generating a transformation plan that, when applied to a sampled source dataset, maximises its Data Quality Score (DQS) and standardises it according to the target dataset. The DQS is a custom metric ranging from 0.0 (poor quality) to 1.0 (high quality), computed by averaging the ratios of missing values, numerical outliers, and duplicated rows. In particular, let $n$ be the total number of non-header cells in the IR, and $m$ be the number of non-header rows. We define the quality indicators as follows:
\begin{itemize}
    \item $M = \text{number of missing entries} \div n$
    \item $O = \text{number of outliers} \div n$
    \item $D = \text{number of duplicated rows} \div m$
\end{itemize}
The overall DQS is then computed as:
\[DQS(\text{IR}) = 1 - (M + O + D) \div 3\]

A higher DQS indicates better data quality, with 1.0 representing a perfectly clean dataset. The transformation plan is structured as a Directed Acyclic Graph (DAG) composed of Data Task Nodes (DTNs), each representing a specific transformation step. These steps iteratively enhance data quality by resolving structural and content-related issues in the source file, aligning it with the format of the target file.

\subsection{Planning Engine - Data Task Nodes}

Each Data Task Node (DTN) receives a set of inputs which include an Internal Representation (IR) along with any additional arguments specific to the transformation logic. The node then applies its processing logic to the IR and passes the updated output to subsequent nodes in the transformation plan. This input-output consistency allows DTNs to be composed into sequential data wrangling plans, facilitating the seamless integration of additional nodes in the future. Each node may optionally accept a strategy parameter, which determines how it addresses its specific task. The current set of DTNs in the FlowETL ecosystem includes the Missing Value Handler (MVH), the Duplicate Rows Handler (DRH), and the Numerical Outliers Handler (NOH). Note that, although our current system focuses on these three tasks plus data standardisation (canonicalization) implemented by Schema Matching in the following subsections, in our future work we could easily plug in other data quality tasks such as outliers of non-numerical data (anomalies), and non-stationary data detection~\cite{nazabal}.

\subsubsection{Missing Value Handling}

The MVH node is responsible for detecting and handling missing values within the IR. Its input parameters are the IR to be operated on, its headers schema, and a third \texttt{strategy} parameter which determines whether the missing values will be imputed with an appropriate placeholder for their data type or dropped, either by column or row. This node has been designed to detect missing values as either empty cells or cells marked with the implementation language's null representation.

\subsubsection{Duplicate Rows Handling}

The DRH node identifies and removes duplicate rows, which are generally redundant for most data analysis tasks~\cite{furche}. This node does not require a \texttt{strategy} parameter, since all duplicated rows are inherently dropped. In this context, rows are considered as singular objects, meaning that rows are equal if they contain the same values for the same columns, disregarding column ordering to account for possible column shuffling during the inward and outward translation mechanism from source file to IR and vice versa. The de-duplication procedure is hash-based. Each row is hashed into a string representation, which becomes the key in the underlying lookup table. If two rows $A$ and $B$ are identical, their hashes will collide, signalling that $B$ is a duplicate of $A$. Once the algorithm terminates, the map's keys represent the unique rows of the IR, and de-hashing them restores the IR without duplicate rows.

\subsubsection{Numerical Outliers Handling}

The NOH node was designed to address outliers in numerical columns only. This design choice was made to offer a foundational yet effective solution to address this data engineering task. Consequently, FlowETL does not currently account for outliers in non-numerical columns. The Median Absolute Deviation (MAD), a statistical method for numerical outlier detection, was chosen due to the median's robustness against outliers, compared to the mean~\cite{yang}. Other approaches such as machine learning-based methods were also considered. Ultimately, MAD was selected due to its simplicity and robustness. The MAD score for a particular point is calculated as:
\[T_{\text{min}} = \text{median}(X) - a \cdot \text{MAD}\]
\[T_{\text{max}} = \text{median}(X) + a \cdot \text{MAD}\]
\[\text{MAD} = b \cdot \text{median} \left( |X - \text{median}(X)| \right)\]

where median and MAD are the corresponding statistics of the outlier scores, the values of $a$ and $b$ are suggested to be 1.48 and 3.0 respectively by Singh et al.~\cite{singhOutliers}, and $X$ represents the $n$ original numerical values. Here, $T_{\text{min}}$ and $T_{\text{max}}$ establish a threshold boundary for normal data points, and observations outside this range are considered to be significantly different from the majority of the data, labelling them as outliers.

In summary, the NOH node first determines the data type of each column. If the column is of type \texttt{number}, an outlier mask of the column is generated by applying statistical methods to detect anomalous values. Flagged outliers are then handled by either imputing the outlier values with the median of the column, or removing the entire row containing the outlier, depending on the value of the \texttt{strategy} parameter.

\subsection{Planning Engine - Schema Inference}

The schema inference step is fundamental to the entire planning process, given that most operations are applied differently according to the type of the column or cell acted upon. Two approaches were considered for schema inference. The first one leverages Apache Spark's \texttt{inferSchema} keyword attribute within its \texttt{read()} method, while the second approach involved defining a custom schema inference method, described by \cref{alg:infer-schema}. The latter was chosen for its flexibility in customising inference logic to suit FlowETL's internal data types and use case. Two major assumptions were made during the design of the schema inference algorithm. First, any cell that can be parsed as a floating-point number is assumed to represent a numeric value. This led to misclassification in columns containing boolean-like data, such as 0s and 1s, which were incorrectly treated as numerical. To mitigate this, a second assumption was introduced: if a column contains exactly two distinct values, it is likely to represent a boolean attribute. This approach effectively captures various binary encodings (e.g., "Y/N", "true/false"), but introduces the risk of misclassification when a column contains only two values due to sampling within the Source Observer.

\begin{algorithm}[t]
\caption{Internal Representation Schema Inference}
\label{alg:infer-schema}
\footnotesize
\begin{algorithmic}[1]
\State \textbf{Input:} \texttt{IR} : Internal Representation
\State \textbf{Output:} \texttt{schema}
\State \texttt{schema} $\gets$ \{\}
\For{\texttt{column in IR.headers}}
    \State \texttt{typeCounts} $\gets$ \{\}
    \State \texttt{valueCounts} $\gets$ \{\}
    \ForAll{\texttt{cell in column}}
        \If{\texttt{cell is null or empty}} 
            \State \textbf{continue}
        \EndIf
        \State \texttt{type} $\gets$ \textproc{inferCellType}(cell)
        \State \texttt{hash} $\gets$ \textproc{serialise}(cell)
        \State \texttt{typeCounts[type]} $\gets$ \texttt{typeCounts[type] + 1}
        \State \texttt{valueCounts[hash]} $\gets$ \texttt{valueCounts[hash] + 1}
    \EndFor
    \If{\texttt{valueCounts.length = 2}}
        \State \texttt{inferredType} $\gets$ \texttt{"boolean"}
    \Else
        \State \texttt{recordedTypes} $\gets$ \texttt{typeCounts.keys}
        \If{\texttt{recordedTypes.length > 1}}
            \State \texttt{inferredType} $\gets$ \texttt{"ambiguous"}
        \Else
            \State \texttt{inferredType} $\gets$ \texttt{recordedTypes[0]}
        \EndIf
    \EndIf
    \State \texttt{schema[column]} $\gets$ \texttt{inferredType}
\EndFor
\State \Return \texttt{schema}
\end{algorithmic}
\end{algorithm}

The schema inference process iterates over all non-missing values, as they are assumed to provide no relevant information on their column's type. For each value, the inferred type and the cell's hashed value are collected. Once the column has been processed, the schema inference algorithm uses the frequency of inferred types and the count of distinct values to determine the column's type. For instance, if multiple types are detected (e.g., \texttt{[number, number, complex]}), the column is considered ambiguous. If exactly two distinct values are found, it is likely a boolean column, as per the earlier assumption. This process is repeated for every column to generate the final output schema for the IR.

\subsection{Planning Engine - Schema Matching}

The main requirements for a schema matching solution are to support one-to-one, many-to-one, and one-to-many matches, while leveraging both syntactic and semantic similarities between columns in the source and target schemas. This step relies exclusively on column names and types, ignoring the actual column values. This design choice reflects the fact that values, types, or structures within matching columns may be altered by subsequent transformation logic, so schema matching must remain agnostic to such changes. Therefore, schema matching is performed prior to transformation, producing an intermediate Internal Representation (IR) that facilitates the later large language model (LLM) inference process. Three different approaches were evaluated as potential solutions for this schema matching step.

The first solution considered, SMUTF~\cite{smutf}, uses XGBoost and sentence transformers, achieving an average F1 score of 0.77 during its evaluation stage. However, its limited training data, lack of control over matching, and inability to handle many-to-one matches made it unsuitable for the given requirements.

The second approach uses an incremental, heuristic-driven method to generate multiple candidate schema matches rather than a single mapping~\cite{gal}. It models schema matching as a bipartite graph problem, assigning similarity scores to all possible element pairs and iteratively refining the top-$K$ mappings by discarding low-scoring edges and applying \emph{Stability Analysis} to prioritise frequently occurring matches. Although this method supports one-to-many mappings, it does not handle many-to-one mappings, limiting its applicability for the task.

The third solution leverages a Large Language Model (LLM) guided by a structured prompt containing a task description, a set of rules to follow when computing a schema matching, matching constraints, and the two schemas.

\subsection{Planning Engine - Plan Evaluation}

The plan evaluation step exhaustively computes and evaluates all possible transformation plans to identify the one that maximises the Data Quality Score (DQS), leveraging the artifacts available to the Planning Engine. The current implementation employs a brute-force strategy, systematically exploring all valid permutations of node-strategy combinations. While computationally intensive, this approach guarantees comprehensive evaluation. FlowETL currently supports three distinct Data Task Nodes (DTNs) and six node-strategy variants, yielding a total of 36 possible plans under the constraint that one node-strategy pair is selected from each category.

The Planning Engine applies each candidate plan to a copy of the sampled Internal Representation (IR) and computes the resulting DQS, selecting the one yielding the highest score. To mitigate computational overhead, the search is terminated early if a plan achieves a DQS above 0.95, under the assumption that additional improvements may not justify the cost of continued exploration. If no valid plan is found, a default non-failing plan is returned.

Certain plans are inherently prone to failure; for example, executing outlier detection prior to missing value imputation can lead to errors, as the outlier handler does not support null values. While not universally optimal, this fallback ensures pipeline continuity and prevents downstream failures.

\subsection{Planning Engine - Inference of Transformation Instructions}

Given possibly multiple source columns $x1, x2, \dots \in X$ a corresponding target column $y$, this component aims to infer a transformation function $f(x,\dots) \rightarrow y$ which combines the inputs to obtain the output column. A key constraint in this process is the exclusion of any external auxiliary data, such as dictionaries, databases, or knowledge bases. As a result, the transformation logic can be inferred solely from the available schema mappings, the structure of the source and target tables, and values within the involved columns.

Several approaches were considered for this step, including Transform-by-Pattern~\cite{jin} and Transform-by-Example~\cite{heTBE}. The chosen method utilises Large Language Models (LLMs) for code generation, guided by a structured prompt containing relevant artifacts to produce a function of the desired form. Zero-Shot Learning was chosen as the prompting technique, enabling the model to perform the task without explicit training on similar examples by leveraging its generalisation capabilities and prior knowledge.

\subsection{Planning Engine - Payload Construction}

The final step in the Planning Engine's workflow is the publishing phase, during which both the computed transformation plan and the Planning Engine's runtime metrics are sent to the Messaging System. This approach relies on the assumption that a transformation plan computed for a representative sample of the source file generalises effectively to the entire source dataset. The plan is structured as follows:

\begin{itemize}
    \setlength{\itemsep}{0pt}  
    \item \texttt{Source File} - name of the file targeted by the plan.
    \item \texttt{Reconstruction Key} - an optional key used to convert the Internal Representation back into JSON format.
    \item \texttt{Schema Map} - a source-target columns mapping as defined by the target file.
    \item \texttt{Plan Steps} - a sequence of DTNs designed to maximise the data quality (DQS) on the sample IR.
    \item \texttt{Logic} - A string containing valid executable instructions generated by the LLM, which is compiled into a Python function and invoked onto the IR by the ETL Worker.
    \item \texttt{IR Schema} - The schema inferred from the sample IR, stored to prevent redundant re-computation within the ETL Worker.
\end{itemize}

\subsection{ETL Worker}

The ETL Worker is responsible for executing the transformation plan generated by the Planning Engine, following the standard ETL methodology of extraction, transformation, and loading. Each worker begins by extracting the file name and associated transformation instructions from the published plan. The source file is then translated into an Internal Representation (IR), upon which a pre-transformation Data Quality Score (DQS) is computed. Next, the worker parses and applies the transformation plan to the IR, followed by the computation of a post-transformation DQS. Finally, the transformed IR is converted back into its original file format and loaded to the designated destination. Decoupling the plan generation from its application allows multiple ETL Workers to be instantiated. The Planning Engine computes the plan once, which can then be distributed to any number of ETL Worker instances, enabling scalable processing based on the size of the source file.

\subsection{Reporting Engine}

The Reporting Engine offers an interface for monitoring the status of individual pipeline components. It continuously polls the Messaging System for runtime metrics and autonomously updates itself with payloads from each component, resulting in a self-managing and self-populating report. By automatically generating reports according to a predefined template, the Reporting Engine ensures consistent output. This structured approach enables users to extend system capabilities for further analysis of the compiled reports. For instance, key metrics could be used to identify performance bottlenecks or anomalies within the pipeline's behaviour; a strategy widely adopted and supported in the literature~\cite{ali, mondal}.

\section{Experimental Results}\label{sec:evaluation}

In this section we present the experimental methodology adopted to evaluate FlowETL. The intrinsic component of the evaluation focuses on the Planning Engine, comparing two different versions, and exploring how altering the sampling percentage within the Source Observer changes the final output. The extrinsic component looks at comparing FlowETL with another ETL tool with respect to output quality. The result are reported and analysed, and limitations of the experimental design are also discussed.

\subsection{Methodology}

The evaluation corpus was constructed by collecting 13 datasets (7 in CSV format, 6 in JSON format) from Kaggle.com and one CSV dataset provided by the University of Aberdeen Chemistry Department (\texttt{Chemistry Field Readings}), ensuring diversity with respect to their domain and size (i.e., entries count). For each dataset, a human-defined ground truth (GT) was created to contain a diverse set of schema-mapping requirements and transformations, including merging, formatting, and formula application. Additionally, a GT transformation plan was defined for each dataset, representing the correct sequence of operations required to convert the source into its target form. The datasets sourced from Kaggle.com contained an insufficient number of data wrangling issues, due to the fact that Kaggle services pre-cleaned datasets ready for downstream tasks. To account for this shortcoming, a polluter script was developed and executed on each dataset, effectively transforming each dataset to contain around 40\% missing values, 20\% duplicated rows, and 5-10\% numerical outliers.

To evaluate the Planning Engine, the \emph{PlanEval} metric was designed. The core idea of behind this metric is inspired by another metric, Success Rate for Data Transformations (SRDT)~\cite{jinsdrt} which aims to capture the percentage of correctly generated transformations. Another inspiration for \emph{PlanEval} comes from BIRD~\cite{li}, a testing framework which focuses on evaluating how well LLMs can solve Text-to-SQL tasks. Within the BIRD framework, the authors define a Valid Efficiency Score (VES) metric, which optimistically computes the correctness of the generated SQL query against the ground truth. Optimism in this context refers to positively scoring an output if and only if it matches the ground truth perfectly, without penalising any incorrect plan components.

\emph{PlanEval} focuses exclusively on evaluating the SRDT, as the associated LLM token usage and API call costs are kept fixed and minimal through a consistent sampling strategy and a fixed planning engine architecture, limiting LLM invocation to exactly two calls per pipeline execution. Additionally, constraining the input to a maximum of 50 Internal Representation (IR) rows per prompt ensures that token consumption, and thus cost, remains within a predictable and controlled range. In this evaluation's context, A plan $P_f = \{m1,m2,...mN,t1,t2,...tM\}$ produced for a particular source file $f$ consists of schema matching steps denoted by $m$ and transformation steps denoted by $t$, acting at the structural and instance level respectively. A ground truth plan $GT_f$ consists of the set of matching and transformation operations required to transform the source file into the target one. The evaluation metric $PlanEval(P_f, GT_f) \rightarrow [0,1]$ rewards correct operations within a plan and ignores incorrect ones, producing a normalised score ranging from 0.0 (completely incorrect plan) to 1.0 (perfect plan).

This metric additionally handles cases where the number of operations in $P_f$ differs from the number of operations in $GT_f$. In particular, if $P_f$ contains missing operations, these are implicitly scored with a 0.0. Similarly in the case where $P_f$ contains extra operations, which might have been hallucinated by the language model, they are also scored with 0.0, making \emph{PlanEval} an optimistic scoring metric. In detail, the score for a plan is computed as follows. First, the maximum achievable score $maxS$ for $P_f$ is initialised as $1.0 \times n$, where $n$ is the number of operations in $GT_f$. Then, the initial score for $P_f$ is set as $s = 0.0$. For each operation $op \in P_f$, if $op \in GT_f$ and $op$ is correct, $s$ is incremented by 1.0; if $op \in GT_f$ but is not correct,  $s$ is incremented by 0.5; otherwise, it is not increment. Finally, the \emph{PlanEval} score for $P_f$ is computed as $s \div maxS$.

\subsection{Intrinsic Evaluation - Algorithmic vs LLM-based Schema Matching}

\begin{figure}[t]
    \centering
    \includegraphics[scale=0.4]{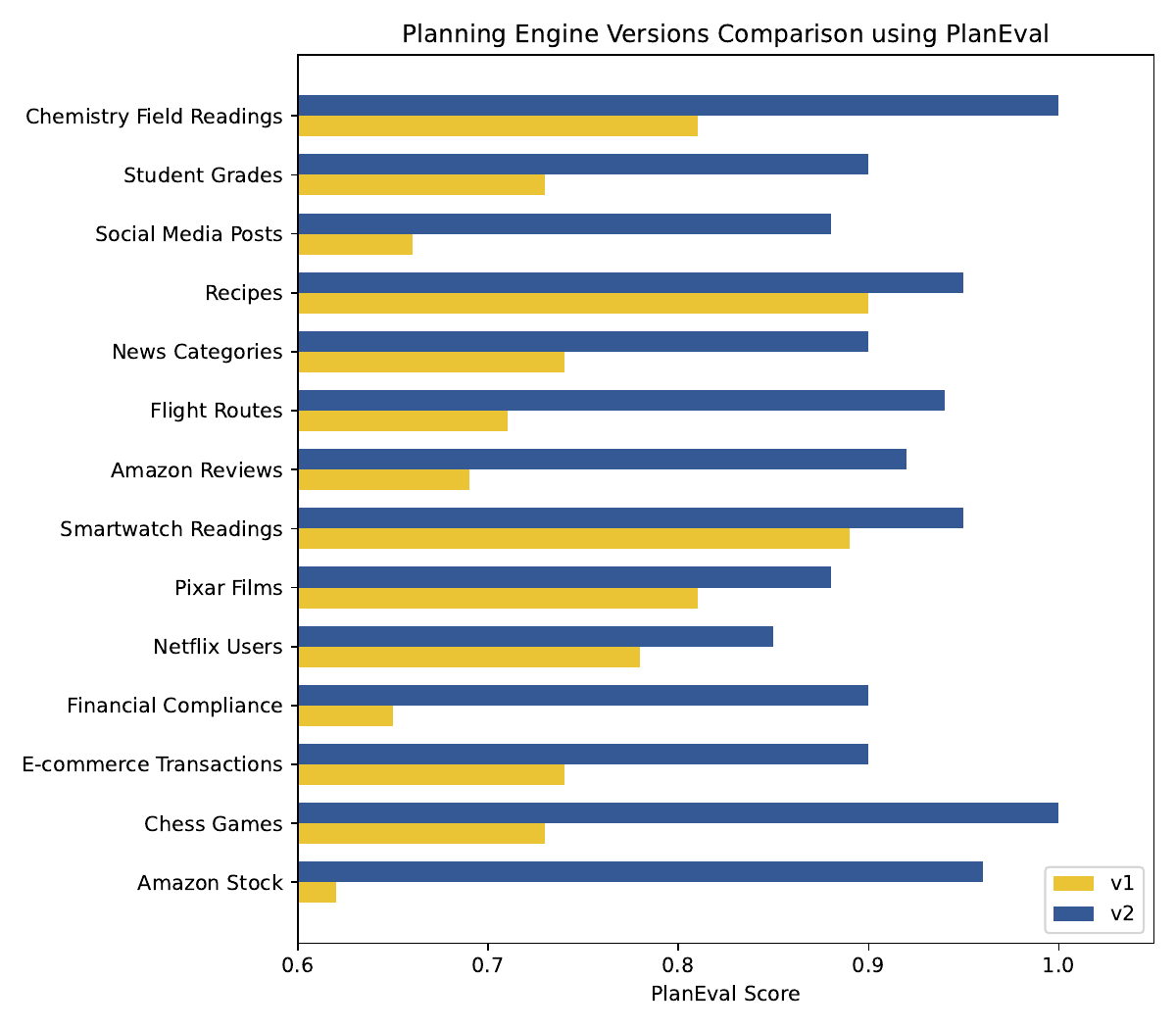}
    \caption{\emph{PlanEval} results from the comparison between the Planning Engine using algorithmic schema matching and example-based prompting (v1, blue) vs the Planning Engine using LLMs for both schema matching and transformation inference, without example-based prompting (v2, orange)}
    \label{fig:v1-v2-figure}
\end{figure}

This experiment compares two versions of the Planning Engine. The first version (v1) achieves the schema matching step through the use of a custom implementation of the Gale-Shapley algorithm~\cite{gale}, extended to support one-to-many and many-to-one matches, as well as providing the large language model (LLM) with a detailed example within the prompt detailing how the transformation logic inference step should be carried out. The second version (v2), uses the same LLM for both steps, approaching the two tasks separately. The transformation inference prompt differs such that the example has been replaced by a set of rules and restrictions to guide the LLM as outlined in Anthropic's documentation to prompting their Claude-3.7-Sonnet model. For each dataset, both versions of the Planning Engine were used to output a plan, which was then scored using \emph{PlanEval}. The results are shown in \cref{fig:v1-v2-figure} and \cref{tab:v1-v2-table}.

Although leveraging an LLM for schema matching introduces a fixed monetary cost, the consistent improvements in plan quality and correctness justify its use. The observed performance gains are likely attributable to the LLM's extensive pre-training on diverse datasets, enabling it to accurately infer mappings, including ambiguous ones. Furthermore, refinements to the prompt structure likely enhanced the model's effectiveness. Additionally, the experiment highlights a limitation of the algorithmic approach used for schema matching, namely that edge weights in the bipartite graph are derived from averaged semantic and syntactic similarities, which leads to poor generalisation in absence of context~\cite{zhou}.

\begin{table}[t]
    \centering
    \caption{Comparison of Planning Engine \emph{PlanEval} score using algorithmic schema matching and example-based prompting (v1) vs LLM-driven schema matching and transformation inference (v2).}
    \label{tab:v1-v2-table}
    \begin{tabular}{| l | c | c |}
    \hline
    \textbf{Dataset} & \textbf{v1 Score} & \textbf{v2 Score} \\
    \hline
    Amazon Stock & 0.62 & 0.96 \\
    \hline
    Chess Games & 0.73 & 1.0 \\
    \hline
    E-commerce & 0.74 & 0.90 \\
    \hline
    Financial Compliance & 0.65 & 0.90 \\
    \hline
    Netflix Users & 0.78 & 0.85 \\
    \hline
    Pixar Films & 0.81 & 0.88 \\
    \hline
    Smartwatch Readings & 0.89 & 0.95 \\
    \hline
    Amazon Reviews & 0.69 & 0.92 \\
    \hline
    Flight Routes & 0.71 & 0.94 \\
    \hline
    News Categories & 0.74 & 0.90 \\
    \hline
    Recipes & 0.90 & 0.95 \\
    \hline
    Social Media Posts & 0.66 & 0.88 \\
    \hline
    Student Grades & 0.73 & 0.90 \\
    \hline
    Chemistry Field Readings & 0.81 & 1.0 \\
    \hline
    \end{tabular}
\end{table}

\subsection{Intrinsic Evaluation - Sampling Percentage}

This experiment evaluates whether varying the sampling percentage within the Source Observer impacts the output of the Planning Engine. It was hypothesised that as the sampling percentage increases, the execution time of the Planning Engine would also increase, however the likelihood of errors or failures in the computed plan would decrease. This is because a larger sample should provide more information for the LLMs to process, but the time required to pass the sample between data task nodes would inherently slow down the planning process.

The experiment was carried out as follows. Firstly, a series of \texttt{p} values from 5\% to \%100, with an increasing step of 5\% was chosen. To select a dataset for this experiment, the median (around 7000) number of objects/rows across all 14 datasets was computed, and the dataset with the closest object count, namely \texttt{Amazon Stock} was selected. For each sampling value \texttt{p}, the time elapsed from the beginning to completion of the Planning Engine, the plan quality scored using \emph{PlanEval}, and the maximum Data Quality Score (DQS) achieved on each sample were recorded. The results are summarised in \cref{tab:sample-experiment-result}.

\begin{table}[t]
    \centering
    \caption{Results of increasing the sampling percentage \texttt{p}.}
    \label{tab:sample-experiment-result}
    \begin{tabular}{| c | c | c | c |}
    \hline
    \textbf{Sample \%} & \textbf{Time Elapsed (s)} & \textbf{PlanEval Score} & \textbf{Max DQS} \\
    \hline
    5  & 69 & 0.85 & 0.98 \\
    \hline
    10 & 66 & 0.85 & 0.96 \\
    \hline
    15 & 78 & 0.90 & 0.97 \\
    \hline
    20 & 70 & 0.90 & 0.96 \\
    \hline
    25 & 65 & 0.85 & 0.95 \\
    \hline
    30 & 73 & 0.90 & 0.96 \\
    \hline
    35 & 71 & 0.85 & 0.96 \\
    \hline
    40 & 78 & 0.95 & 0.95 \\
    \hline
    45 & 75 & 0.90 & 0.96 \\
    \hline
    50 & 81 & 0.95 & 0.96 \\
    \hline
    \end{tabular}
\end{table}

\begin{figure*}[t]
    \centering
    \includegraphics[scale=0.45]{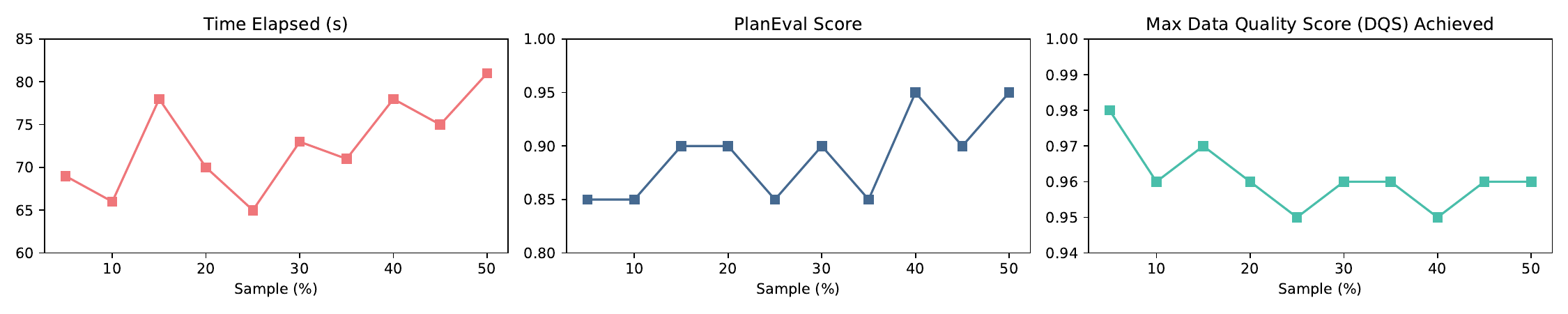}
    \caption{Time elapsed (left), \emph{PlanEval} score (middle), Maximum Data Quality Score (DQS) achieved (right) on the \texttt{Amazon Stock} dataset for varying sampling percentage \texttt{p} values}
    \label{fig:sampling_experiment_chart}
\end{figure*}

The pipeline failed to handle any \texttt{p > 0.5} due to the large payload size. While the results were incomplete, a distinct series of patterns emerged, as shown in \cref{fig:sampling_experiment_chart}. The elapsed time appeared to increase gradually with the sample size. This behaviour was expected, as many data engineering tasks, such as handling missing values and removing duplicates, require a linear scan of the Internal Representation (IR), causing the time taken to grow linearly with the input size. The \emph{PlanEval} score also scaled linearly with \texttt{p}, likely because larger samples provide more representative data, offering the LLM more context to correctly infer schema matches and transformation steps. The maximum DQS achieved remained consistent, indicating that the sampling percentage is not strongly correlated with \texttt{p}. This conclusion is further supported by the Pearson correlation coefficient analysis, reported in \cref{tab:pearson_correlation_sampling_experiment}.

\begin{table}[t]
    \centering
    \caption{Pearson correlation coefficients for the sampling percentage evaluation task.}
    \label{tab:pearson_correlation_sampling_experiment}
    \begin{tabular}{| l | c |}
    \hline
    \textbf{Correlation Pair} & \textbf{Pearson Correlation} \\
    \hline
    Time Elapsed vs Sample \% & 0.62 \\
    \hline
    PlanEval vs Sample \% & 0.77 \\
    \hline
    Max DQS vs Sample \% & -0.48 \\
    \hline
    \end{tabular}
\end{table}

\begin{figure}[t]
    \centering
    \includegraphics[scale=0.45]{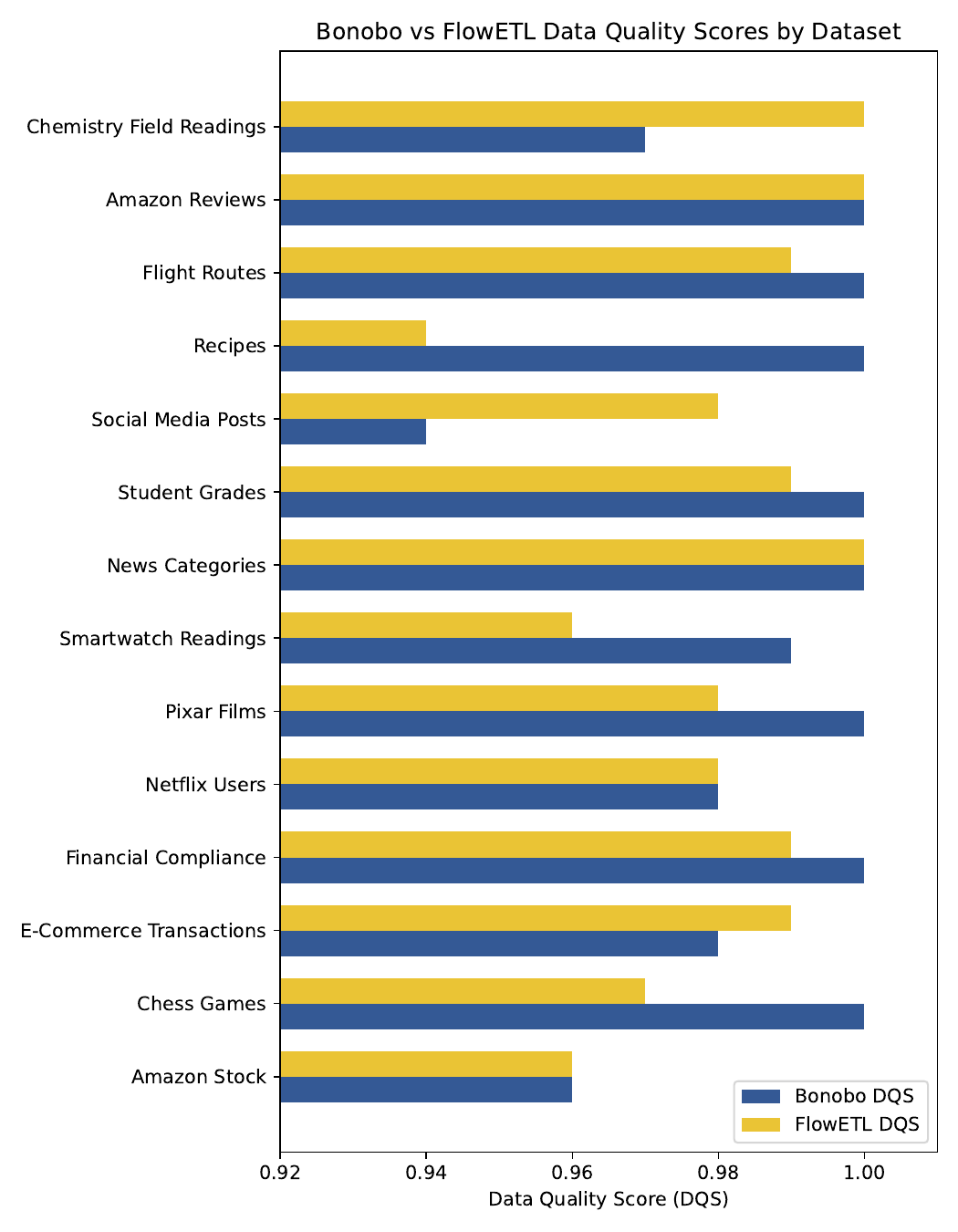}
    \caption{Comparison of the DQS achieved on each evaluation dataset by FlowETL (yellow) and Bonobo (blue)}
    \label{fig:flow-etl-dq-achieved}
\end{figure}

Two major limitations of this experiment's design were the monetary costs associated with running FlowETL, and the Planning Engine's capabilities of only handling smaller IRs which fit within the payload size tolerated by XComs, the Planning Engine's internal messaging system provided by Apache Airflow. This limitation likely contributed to pipeline failures when processing larger datasets. Repeating the experiment with additional datasets, potentially capping \texttt{p} to ensure that the IR stays within the payload limit, could produce a more representative set of results.

\subsection{Evaluation Against Competing Tools}

Another approach to evaluate FlowETL involved comparing it against other ETL tools available on the market which provided similar functionalities. A major challenge identified during the design of this experiment was the lack of open-source, example-based ETL solutions.

Foofah~\cite{foofah} was a promising candidate, however its limited functionality made direct comparison with FlowETL impractical. Bonobo ETL was selected as an alternative. Bonobo is an open-source Python framework designed for creating lightweight, scalable, and maintainable data pipelines using Directed Acyclic Graphs (DAGs) composed of reusable transformation components. A major issue is that Bonobo is non-autonomous, making it challenging to evaluate FlowETL's planning capabilities against other non-autonomous pipeline solutions.

The experiment was setup as follows. The author first learned to use the basic functionalities offered by Bonobo. Subsequently, a custom Bonobo transformer method was defined for each evaluation dataset, following the ground truth (GT) previously defined, resulting in 14 different Bonobo workflows being constructed for this evaluation. The execution time, Data Quality Score (DQS), missing values percentage, duplicate rows percentage, and outliers percentage were recorded for each dataset after running their respective pipeline. The results are reported by \cref{tab:bonobo-results}.

\begin{table*}[t]
    \centering
    \caption{Results gathered after transforming all datasets using Bonobo. The \textbf{Time} column only indicates each pipeline's runtime and does not account for development time. Missing values, duplicate rows, and outliers columns indicate the percentage detected post-ETL in each dataset. Lower percentages correlate with a higher Data Quality Score (DQS).}
    \label{tab:bonobo-results}
    \begin{tabular}{| l | c | c | c | c | c | c |}
    \hline
    \textbf{Dataset} & \textbf{Entries} & \textbf{Time (s)} & \textbf{DQS Achieved} & \textbf{Missing Values \%} & \textbf{Duplicate Rows \%} & \textbf{Outlier Values \%} \\
    \hline
    Amazon Stock & 7557 & 0.16 & 0.96 & 0.00 & 0.13 & 0.00 \\
    \hline
    Chess Games & 24093 & 0.68 & 1.0 & 0.00 & 0.00 & 0.00 \\
    \hline
    E-Commerce & 650151 & 11.32 & 0.98 & 0.00 & 0.06 & 0.00 \\
    \hline
    Financial Compliance & 117 & 0.01 & 1.0 & 0.00 & 0.01 & 0.00 \\
    \hline
    Netflix Users & 29827 & 0.56 & 0.98 & 0.05 & 0.01 & 0.00 \\
    \hline
    Pixar Films & 36 & 0.01 & 1.0 & 0.00 & 0.04 & 0.04 \\
    \hline
    Smartwatch Readings & 12039 & 0.17 & 0.99 & 0.00 & 0.03 & 0.00 \\
    \hline
    News Categories & 280 & 0.01 & 1.0 & 0.00 & 0.01 & 0.00 \\
    \hline
    Student Grades & 5682 & 0.24 & 1.0 & 0.00 & 0.00 & 0.00 \\
    \hline
    Social Media Posts & 94 & 0.01 & 0.94 & 0.00 & 0.02 & 0.21 \\
    \hline
    Recipes & 51361 & 0.63 & 1.0 & 0.00 & 0.00 & 0.00 \\
    \hline
    Flight Routes & 10695 & 0.28 & 1.0 & 0.00 & 0.00 & 0.00 \\
    \hline
    Amazon Reviews & 1948 & 0.06 & 1.0 & 0.00 & 0.00 & 0.00 \\
    \hline
    Chemistry Field Readings & 65536 & 0.71 & 0.97 & 2.28 & 0.00 & 0.45 \\
    \hline
    \end{tabular}
\end{table*}

The current runtime measurements for Bonobo are not fully representative, as they exclude the time spent analysing the input datasets, interpreting the ground truth, and gaining familiarity with the Bonobo framework prior to implementing the required transformations.

A more accurate evaluation would require the recruitment of developers to implement the required transformations for each dataset, repeating the task for both Bonobo and FlowETL. The hypothesis is that FlowETL would prove easier and faster to use, as specifying the target output for a dataset is expected to require less effort than constructing an equivalent workflow using Bonobo. If the manual implementation takes longer than executing the complete ETL process with FlowETL, it would indicate greater efficiency of the latter. This form of human-in-the-loop evaluation was not originally planned and represents an unaddressed limitation. Therefore, the evaluation focused primarily on FlowETL's ability to handle common data wrangling challenges, alongside correctly inferring and applying a transformation plan.

In addition, the data quality of the transformed output and the corresponding \emph{PlanEval} Score were recorded for each execution. For every dataset, a corresponding target output containing 5 to 7 entries was manually constructed, resulting in 14 source-target dataset pairs. Each pair was processed by FlowETL using a sampling rate defined as $p = \max(\texttt{object\_count} \times 0.05, 50)$. The same evaluation metrics used in the Bonobo experiment were collected to enable a dataset-wise comparison of the pipelines' data wrangling performance. The results are presented in \cref{tab:flowetl_experiment}.

\begin{table*}[t]
\centering
\caption{Runtime results gathered by running FlowETL on all evaluation datasets. The \textbf{Time} column measures the runtime for each dataset, end-to-end, using pre-constructed target datasets. Missing values, duplicate rows, and outliers columns indicate the percentage detected post-ETL in each dataset. Lower percentages correlate with a higher Data Quality Score (DQS).}
\label{tab:flowetl_experiment}
\begin{tabular}{| l | c | c | c | c | c | c |}
\hline
\textbf{Dataset} & \textbf{Time (s)} & \textbf{DQS} & \textbf{Missing Values \%} & \textbf{Duplicate Rows \%} & \textbf{Outlier Values \%} & \textbf{PlanEval Score} \\
\hline
Amazon Stock & 140.2 & 0.96 & 0.00 & 3.41 & 0.00 & 0.96 \\
\hline
Chess Games & 102.3 & 0.97 & 0.00 & 0.00 & 0.00 & 1.0 \\
\hline
E-Commerce & 99.4 & 0.99 & 0.00 & 2.73 & 0.00 & 0.90 \\
\hline
Financial Compliance & 111.9 & 0.99 & 0.00 & 0.00 & 0.00 & 0.90 \\
\hline
Netflix Users & 79.5 & 0.98 & 3.76 & 4.58 & 0.00 & 0.85 \\
\hline
Pixar Films & 95.7 & 0.98 & 1.93 & 3.56 & 2.80 & 0.88 \\
\hline
Smartwatch Readings & 111.6 & 0.96 & 0.00 & 0.00 & 0.00 & 0.95 \\
\hline
Amazon Reviews & 79.0 & 1.0 & 0.00 & 1.32 & 0.00 & 0.92 \\
\hline
Flight Routes & 90.1 & 0.99 & 0.00 & 0.00 & 1.51 & 0.94 \\
\hline
News Categories & 76.2 & 0.99 & 0.00 & 1.73 & 0.00 & 0.90 \\
\hline
Recipes & 88.9 & 0.94 & 0.00 & 0.00 & 0.00 & 0.95 \\
\hline
Social Media Posts & 76.1 & 0.98 & 0.00 & 3.44 & 4.62 & 0.88 \\
\hline
Student Grades & 106.1 & 1.0 & 0.00 & 0.00 & 0.00 & 0.90 \\
\hline
Chemistry Field Readings & 70.2 & 1.0 & 0.00 & 0.00 & 0.00 & 0.96 \\
\hline
\end{tabular}
\end{table*}

As the results show, FlowETL achieved post-ETL data quality scores ranging from 0.94 to 1.0, which are comparable to those achieved with Bonobo. However, the overall DQS across all datasets was slightly lower for FlowETL. This can likely be attributed to a higher incidence of unresolved data wrangling issues in the output, as illustrated in \cref{fig:flow-etl-dq-achieved}. A contributing factor may be poor generalisation of the Planning Engine's output on the entire dataset. Despite these limitations, FlowETL demonstrated strong generalisation capabilities, consistently producing high-quality outputs and \emph{PlanEval} scores, while autonomously inferring and executing transformation steps.

\section{Conclusion and Future Work}\label{sec:conclusion}

This work presented FlowETL, a novel and autonomous ETL pipeline capable of inferring and applying data transformation plans by analysing a source and corresponding target dataset. FlowETL was evaluated on 14 diverse datasets, demonstrating robust performance, high data quality retention (DQS between 0.96 and 1.0), and effective generalisation across both structured and unstructured formats. The system significantly reduced manual intervention compared to traditional tools while maintaining consistent execution time through sampling-based planning. 

Despite its strengths, the ongoing cost of using LLM APIs and the absence of support for data enrichment from external sources are the current main limitations of FlowETL. Future work targets the integration of distributed computing frameworks like Apache Spark for improved scalability, supporting custom LLMs and Data Task Nodes (DTNs) for greater flexibility, and introducing machine learning-based strategies for anomaly detection and imputation. Expanding data type coverage, adding configuration file support for customisation, improving schema matching for nested structures, and developing a graphical user interface (GUI) are also considered. Finally, caching mechanisms and support for streaming data and additional file formats beyond CSV and JSON could improve FlowETL's usability and performance in real-world settings.

\bibliographystyle{ACM-Reference-Format}

\end{document}